\documentclass[]{spie}  
\usepackage[]{graphicx}
\usepackage{float}
\usepackage[table]{xcolor}

\title{ALMA Nutator Design and Preliminary Performance} 

\author{Pierre Martin-Cocher\supit{a}, John Ford\supit{b}, Patrick M. Koch\supit{a}, Chih-Wen Ni\supit{c}, Wei-Long Chen\supit{d}, Ming-Tang Chen\supit{a}, Philippe Raffin\supit{a}, Chin-Long Ong\supit{e}, Paul T.P. Ho\supit{a} and Arthur Symmes\supit{b}
\skiplinehalf
\supit{a}Institute of Astronomy and Astrophysics, Academia Sinica.  No.1, Roosevelt Rd, Sec. 4 Taipei 10617, Taiwan. \\
\supit{b}National Radio Astronomy Observatory, Green Bank, WV 24944 (USA), Greenbank, WV, U.S.A.\\
\supit{c}Realscene Technology Corp., 14-1, Lane 272, Dalin Road, Daya District,Taichung City, 42847, Taiwan.\\
\supit{d}Department of Industrial Design, Chaoyang University of Technology, 168 Jifeng E. Rd., Taichung City, Taiwan.\\
\supit{e}COTECH, Inc., 21,Jing 1st Road., Wu Ci, Taichung 43541,Taiwan.\\
}

\authorinfo{Further author information: (Send correspondence to PMC)\\PMC: E-mail: pierre@asiaa.sinica.edu.tw, Telephone: 886 22366-5375}
 
\begin{document} 
  \maketitle 

\begin{abstract}
We report the past two years of collaboration between the different actors on the ALMA nutator. Building on previous developments, the nutator has seen changes in much of the design. A high-modulus carbon fiber structure has been added on the back of the mirror in order to transfer the voice coils forces with less deformation, thus reducing delay problems due to flexing. The controller is now an off-the-shelf National Instrument NI-cRIO, and the amplifier a class D servo drive from Advanced Motion Controls, with high peak power able to drive the coils at 300 Volts DC. The stow mechanism has been totally redesigned to improve on the repeatability and precision of the stow position, which is also the reference for the 26 bits Heidenhain encoders. This also improves on the accuracy of the stow position with wind loading.  Finally, the software, written largely with National Instrument's LabView, has been developed. We will discuss these changes and the preliminary performance achieved to date. 
\end{abstract}



\section{INTRODUCTION}
\label{sec:intro}  

The Atacama Large Millimeter/submillimeter Array, or ALMA project, in northern Chile, is an international telescope project which is being built on a site at 5 km elevation. The site provides
excellent atmospheric transmission in the millimeter and sub-millimeter wavelength ranges. In order to optimize observations, a nutating subreflector able to rapidly switch between two positions on
the sky ("on source" and "off source") is a fine addition. In the case of the ALMA project, the specifications are quite stringent with a "fly time" between two positions of 10 ms, a positioning
precision of 1.3 arcsec Root Mean Square (RMS), and a maximum nutating amplitude of 1080 arcsec.\\
This document will concentrate on the design changes made during the past two years, as outlined in table 1. Previous progress on the ALMA nutator work has been reported in [\cite{alma},\cite{atacama},\cite{nutator}].

\begin{table}[H]
\caption{Recent Design Modifications} 
\label{tab:fonts}
\begin{center}       
\begin{tabular}{|l||l|l|} 
\hline
\rule[-1ex]{0pt}{3.5ex} \bf Item & \bf Problem  & \bf Modification\\
\hline
\hline
\rule[-1ex]{0pt}{3.5ex}  Mirror & deformation & ribbing, strengthening   \\
\hline
\rule[-1ex]{0pt}{3.5ex}  Stow function & stiffness, repeatability & relocation and closed loop actuators  \\
\hline
\rule[-1ex]{0pt}{3.5ex}  Rocker & creep & new fabrication process  \\
\hline
\rule[-1ex]{0pt}{3.5ex}  Controller & flexibility & off-the-shelf NI-cRIO  \\
\hline
\rule[-1ex]{0pt}{3.5ex}  Power amplifier & drive voltage and peak power & switch to class D amplifier  \\
\hline
\rule[-1ex]{0pt}{3.5ex}  Software & unsatisfactory performance & still in development  \\
\hline
\end{tabular}
\end{center}
\end{table}


\section{MECHANICAL DESIGN} 

This section describes the general design of the nutator, with an emphasis on  the modifications that have taken place in the last two years: mirror strengthening and stow mechanism improvements.
The nutator is entirely made of carbon fiber reinforced plastic (CFRP) elements mounted on an aluminum base that doubles as the interface to the antenna quadrupod.
The mirror is rotating around a spring-loaded axis, and is connected to a set of voice coil actuators mounted in a "push-pull" arrangement. The reaction forces to these coils come from a rocker, itself rotating around its own spring-loaded axis.  The rocker,  shown in deep blue in Figure 1, houses the magnet of the voice coil actuators.  The main function of the rocker is to provide the reaction force necessary to cancel forces generated by the mirror motion from the mounting base plate. This is necessary to minimize the residual torque transmitted to the quadrupod assembly. The angular positions of both the mirror and the rocker are measured relative to the base plate by absolute 26-bits encoders.\\
As described, the mirror-rocker arrangement makes for a tuned oscillator, but has no provision for reaction to perturbations outside of this system. In effect, the absolute position of the rocker relative to the base plate is not well defined and subject to outside forces (wind load being the predominant one). To provide some compensation to the outside forces, there is an additional set of smaller voice coils mounted between the rocker and the base plate (yellow in figure 1).

   \begin{figure}[H]
   \begin{center}
   \begin{tabular}{l r}
   \includegraphics[height=6cm]{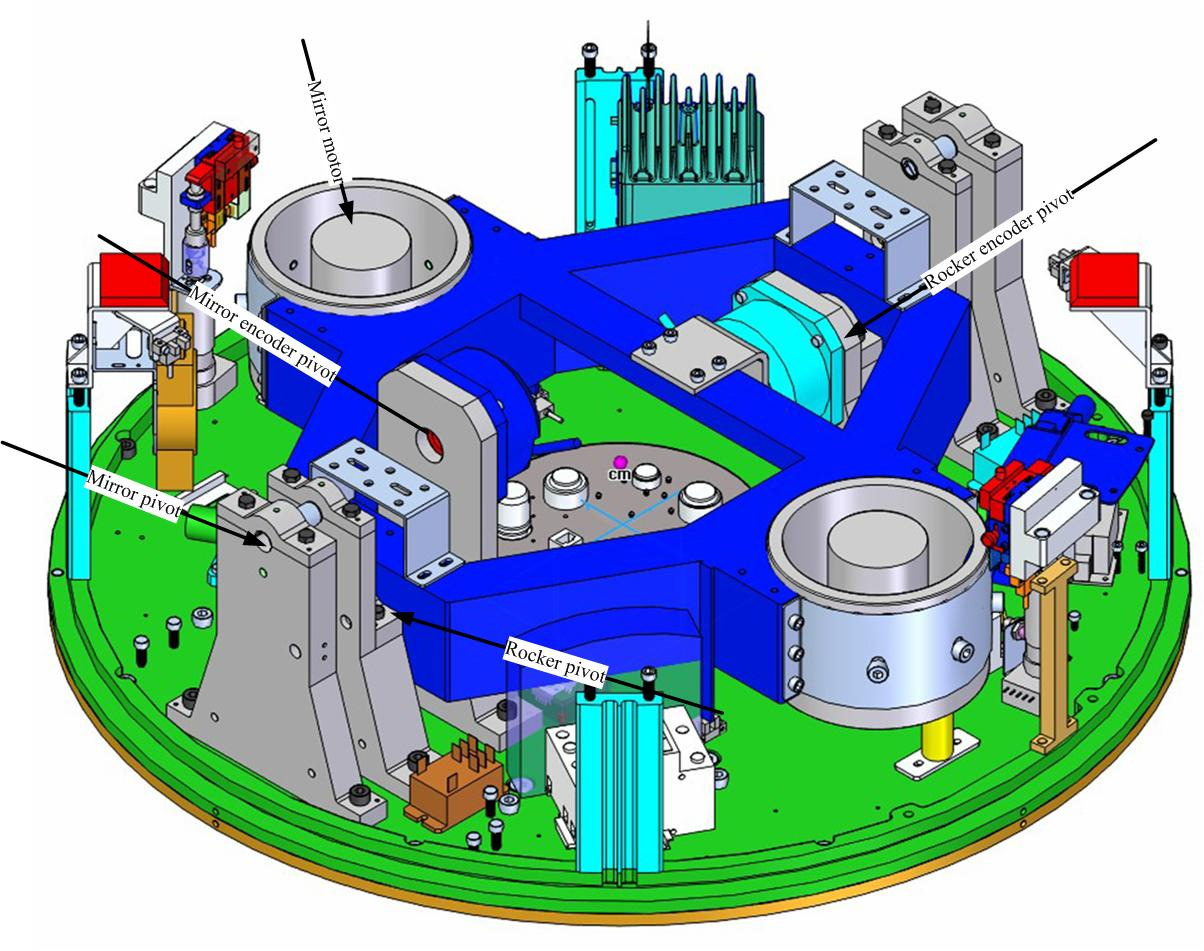} & \includegraphics[height=6cm]{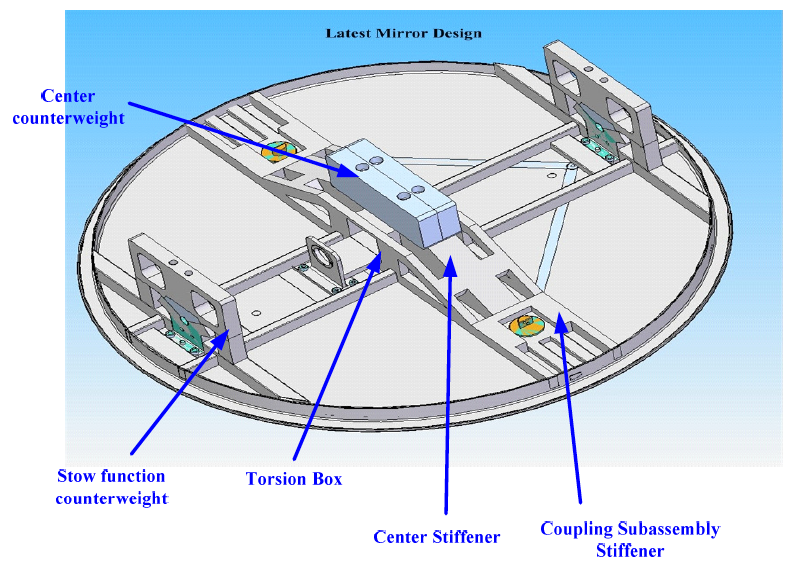} \\
   \end{tabular}
   \end{center}
      \caption[example] 
   { \label{fig:1} 
Left panel: The nutator mechanical assembly, showing the main components, mirror removed. In deep blue, the rocker on which are mounted the voice coil magnets. Right panel: The mirror bottom view showing the additional ribbing. }
   \end{figure}

\subsection{Mirror Strengthening} 
Two years ago, it was found that the mirror, as designed, experienced severe deformations during the transitions. The actuating coils can produce a 50-pound force (lbs) continuously for a 2 A current (limited by the maximum heat dissipation). In our design though, the motors are used at up to ten times that rating, with a low duty cycle. The corresponding forces on the mirror imparted an S-shaped deformation to it, and since the encoder  mechanical attachment is in the middle section, there was an initial 5 to 7 ms delay in the apparent motion of the mirror. This translated into an unacceptable phase shift that was difficult to deal with in the controller. Furthermore, this initial stimulus would excite poorly damped vibratory modes of the mirror, thus further complicating the mirror control. The first step in improving the response was to add stiffening ribs to the back of the mirror, as well as relocating the counterweights at the center of the mirror. \\
Extensive finite element simulations through modal analysis helped in designing the  strengthened mirror backup structure, enabling the structure to use the coil's efforts more effectively. The stiffening structure is made of high-modulus carbon fiber in a pre-impregnated fabric (Fig.~\ref{fig:1}, right panel).

\subsection{Stow Function} 
\begin{figure}[H]
   \begin{center}
   \begin{tabular}{l r}
   \includegraphics[width=8cm]{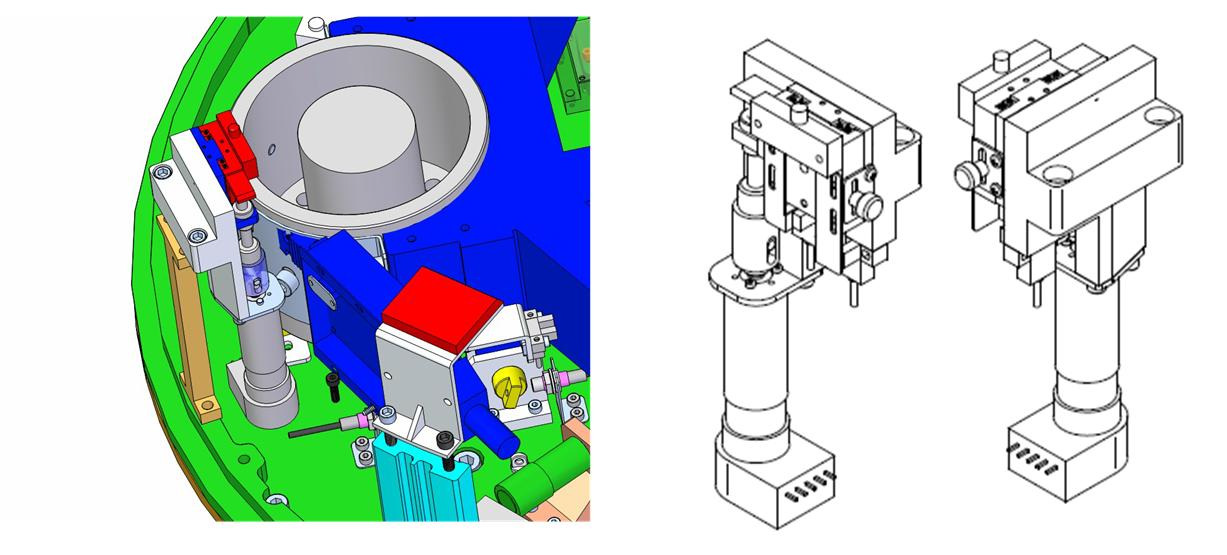} & \includegraphics[width=8cm]{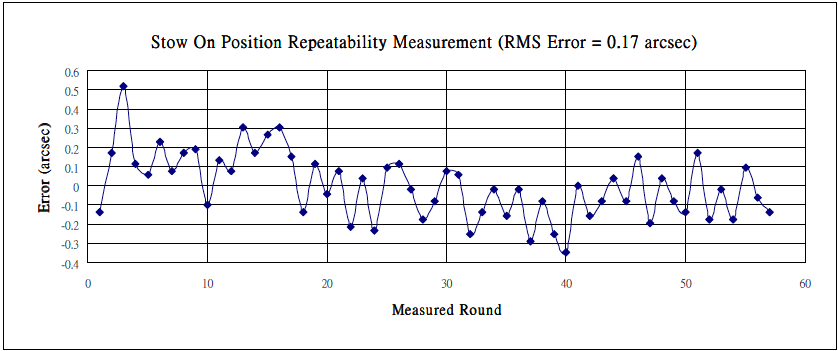} \\
   \end{tabular}
   \end{center}
      \caption[example] 
   { \label{fig:2} 
Left panel: The mirror stow actuators. Right panel: Repeatability of successive stows in open loop mode.}
   \end{figure}

The stow function was initially done by rotating cams that would lock the counterweights at the end of the rotating axis (Fig. 1, right panel). Although the cam locking scheme works well enough with the rocker, it was found that the torque imparted to the mirror by the wind loads was too high to be counteracted on with enough stiffness. Furthermore, the relocation of the counterweights during the mirror strengthening forced a redesign. An additional problem is that the stow actuators must enable a precise positioning in the stowed position, and again the cam system was not repetitive enough to guarantee  a precise, repetitive positioning. We, thus, entirely redesigned this subsystem. It is now done by small actuators that lock at the periphery of the mirror, providing added torque.\\
We were happy at first with the repeatability of the stow assembly (see Fig.~\ref{fig:2}, 57 cycles of free/stow position with and RMS error of 0.17 arcsec), but later found out that in the stowed position, the mirror position would drift with varying temperature and/or wind. We, therefore, are modifying the microcontroller's stow assembly to close the loop on the mirror encoder. In that fashion, the stow position should be only limited by the encoder resolution.

\subsection{Rocker Redesign} 
\begin{figure}[H]
   \begin{center}
   \begin{tabular}{l r}
   \includegraphics[width=8cm]{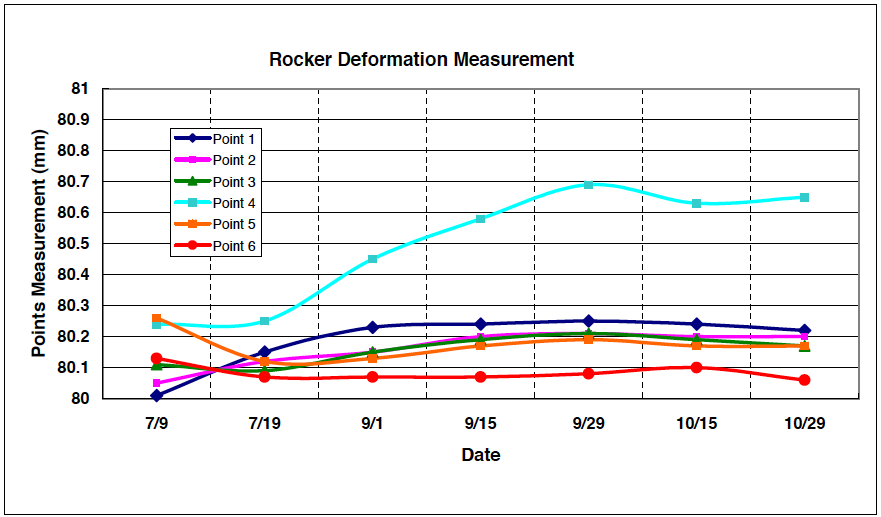} & \includegraphics[width=8cm]{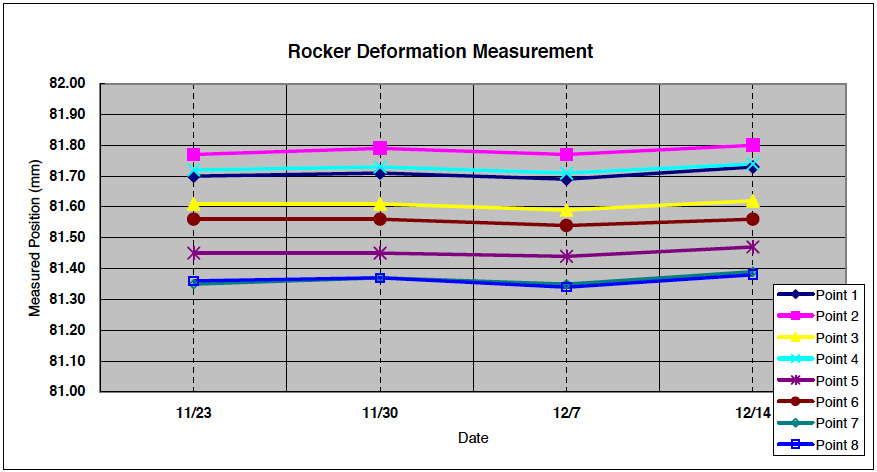} \\
   \end{tabular}
   \end{center}
      \caption[example] 
   { \label{fig:3} 
Left panel: deformations of the original rocker, six points were monitored regularly. Right panel: eight points monitored on the new rocker}
   \end{figure}

One consequence of the mirror strengthening was the relocation of the mirror counterweights. This in turn called for a rocker modification where the existing central rib was moved to the periphery in order to allow clearance for the counterweights.  During that process, it was suspected that the rocker material was creeping. A careful measurement showed that this, indeed, was the case (See Fig.~\ref{fig:3} ). Six (eight for the new rocker) points were monitored around the rockers (next to the rotation axis and the magnets mounting area) while loaded with weights equivalent to the magnets. The original rocker showed important deformation in particular for the first two weeks after manufacturing, while the new rocker is better behaved.\\
The new rocker had been designed using a different process from the original one. The original design called for a dry fiber wrapped around a  polyurethane (PU) core, then infused with epoxy and cured
at room temperature. The new rocker is made of epoxy pre-impregnated high modulus fiber wrapped around a PU core and cured at 120 degrees Celsius. 

\section{ELECTRONIC HARDWARE} \label{sec:sections}

Some of the electronic hardware was found to be inadequate and was, thus, upgraded or changed altogether. Beside improving performances, this redesign was aimed at using properly rated off-the-shelf components for the sake of easier upgrading, follow-up and 
inventory.

\subsection{Controller} 
The initial controller was a dedicated circuit built around a Xilinx Field Programmable Gate Array (FPGA) and custom input/output (I/O) interfaces. Although this solution had advantages in size and weight, it was decided to use a National instrument (NI) cRIO system instead due to its easy programmability. There are two of these NI controllers for each nutating system: one in the nutator itself, used for the basic interlock system, and one in the controller electronics rack that integrates most of the functions. These controllers are programmed using NI's LabView. On a side note, all the housing racks have been changed from in-house to off-the-shelf Electro-Magnetic Interference (EMI) rated units.

\subsection{Power Amplifier} 
The first power amplifier (a linear amplifier) was found to lack sufficient power to move the nutator during the position change, because this transition demands a very high peak force. It has been changed to a class D ("PWM") amplifier, since these are better suited for high-peak currents. Another advantage of the PWM amplifier in this application is that, compared to a linear amplifier, the dissipated power in a PWM amplifier is at a minimum, thus reducing the cooling requirement, which is always a challenge at high elevations. 

During development, it was found that ideally one should use two amplifiers: a powerful one with a high peak rating for the transition time, and a low gain, low noise amplifier for the steady part of the motion where the objective is to keep the mirror at a constant angle with the required precision (1.3 arcsec RMS). This was achieved by switching the amplifier transconductance from a low value (for the steady-state) to the maximum value of 4.2 A/V using a shunt transistor across the gain setting potentiometer, the gate of this transistor being connected to one of the controller digital outputs through an optical coupler.
The Advanced Motion Control class D amplifier model B30A40 can drive the coils at 300 V nominal (up to 400 V with the back EMF), and at a maximum current of 30 A peak (15 A  continuous).

\section{SOFTWARE CONTROL}

The software for the system is written with the Labview programming environment. The servo is
done with the FPGA module, while the supervisory controls and CAN bus interactions are handled
by the real-time computer in the cRIO base module. All of the servo code is implemented in fixed-point arithmetic. The timing of the control loop is controlled by an internal timer. The timer
starts the encoder data acquisition process, and when that completes, the servo loop is run. Upon
completion of the servo loop calculations, the D/A converter is updated with the new command.
The system then waits for the next timer event.

\begin{figure}[h]
   \begin{center}
   \begin{tabular}{c}
   \includegraphics[height=6cm]{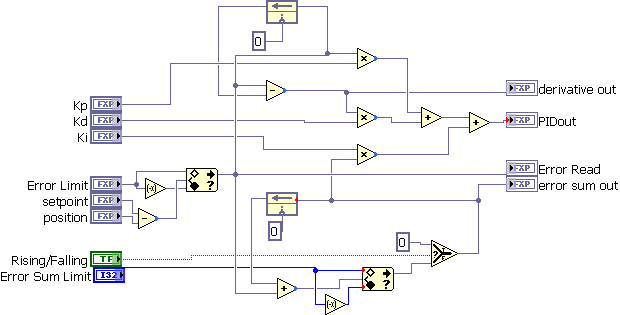}
   \end{tabular}
   \end{center}
   \caption[PID] 
   { \label{fig:PID} 
The implementation of the PID controller in the FPGA.}
   \end{figure} 

\subsection{Implementation Considerations}  
Implementing the servo has been done in the FPGA, using the labview tools (See Fig. 4 for a PID implementation). This results in a very
fast loop rate, but limited arithmetic range and accuracy due to the fixed-point numbers. In addition, the
delays in the I/O subsystem were not really well-known. 
After testing, it was found
that the servo loop requires 
$64$ microseconds to just do the I/O. Calculation time
is a few microseconds, since it is done in parallel inside the FPGA. A loop time of at least 100
microseconds will allow the I/O to complete. A 50-microsecond loop will not.
The loop is arranged thus: Read encoder $\rightarrow$ calculate $\rightarrow$ write analog outputs $\rightarrow$ wait until next loop timer event.

\subsection{Nutator Dynamic Model} 
The full nutator model consists of coupled rocker and mirror models. In this section, we  have chosen to work exclusively with the mirror model so that the effects of the mirror servo can be modeled alone. The rocker servo bandwidth is low enough that the mirror position will not be affected by the rocker servo.
From the Realscene model, the plant transfer function Q(s) is as follows, where N(s) and D(s) are
the numerator and denominator polynomials:

\begin{equation}
N(s) = 1.5873s^{2} + 0.8540s + 116.6070,
  \label{eq1}
\end{equation}

\begin{equation}
D(s) = 1.0000s^{4} + 53.9155s^{3} + 64343.8054s^{2} + 320581.0359s + 4760885.4642,
  \label{eq2}
\end{equation}

\begin{equation}
Q(s) = \frac{N(s)}{D(s)}.
  \label{eq3}
\end{equation}

Since the amplifier closes the current loop, the motor and amplifer combination is modeled as a
series combination of the 5 kHz low-pass characteristic of the amplifier, and the 107 Nm/A torque
constant of the motor. This transfer function is given by:

\begin{equation}
M(s) = (4.2A/V)(107.85Nm/A)(\frac{1}{3.1831e^{-5}s + 1}).
  \label{eq4}
\end{equation}

The series combination of the motor, amplifier, and plant comprise the open-loop transfer function
of the nutator, given by:

\begin{equation}
P(s) = {Q(s)} \times {M(s)},
  \label{eq5}
\end{equation}

or

\begin{equation}
P(s) = \frac{717.7s^{2} + 386.1s + 5.272\times 10^{4})}{(3.183\times 10^{-5} s^{5} + 1.002 s^{4} + 55.96 s^{3} + 6.435 \times 10^{4} s^{2} + 3.207\times 10^{5} s + 4.761\times 10^{6}}. \
  \label{eq6}
\end{equation}

These transfer functions are used with the help of Matlab and Simulink  to analyze and design control algorithms.

\subsection{Control System Design Parameters} 
Digital control system design can provide much freedom in choosing a control law to satisfy the requirements of the system.  Much of the difficulty in designing a control system is in translating the requirements into parameters that can be designed into a system. For instance, systems can be approximately described by simple measures such as damping ratio or settling time. These parameters can then be used to design a control algorithm with these characteristics, which can then be simulated using a tool such as Simulink to control the design.

Most standard control system formulations assume a second-order plant, or one that can be reduced
to second-order. Our plant is 5th order. A reduced-order plant model may be found by factoring
the denominator of the plant transfer function, and separating out the dominant roots into a 2nd
order transfer function. The remaining roots are then replaced by an equivalent DC gain. We found
that 2nd order models did not represent the response very well, so the approximations are not really
valid. This may explain why we had such a hard time trying to tune the PID controller.

\subsubsection{Settling Time}
For our system, settling time is defined as the time for the signal to be
within 1.3 arcsec of its final value. The total step size, in the
worst case, is 1080 arc-seconds which corresponds to an 18 arcmin throw angle. Therefore, 1.3
arcsec amounts to 0.12\% of the step size. We want this to settle in
less than 10 milliseconds. For a second-order system, or a system with 2
dominant roots, the settling time is related
to the damping ratio and the natural frequency of the system by the relation:
$exp (-\zeta\omega_{n} T_{s}) < 0.0012$,
where $T_{s}$ is the settling time.
Taking the logarithm of both sides, we have:

\begin{equation}
\zeta\omega_{n} T_{s} \approx 6.725.
  \label{eq7}
\end{equation}

In other words, 6.7 time constants of the dominant roots of the
characteristic equation. This tells
us that the time constant of the dominant roots must be less than 0.010
sec / 6.7.
We will attempt to design a control system to satisfy the above requirements.

\subsubsection{Percent Overshoot and Damping Ratio} 
In order for the mirror and rocker to stay within the travel limits and remain active, the overshoot must be limited. This implies that the damping ratio (the ratio between the system actual damping and the critical - or optimal - damping)  must be higher. In order for the percent overshoot to be
less than about $1\%$, the damping ratio must be $\approx$ 0.82. \cite{Dorf}

\subsubsection{System Bandwidth} 
Given the damping ratio from 4.3.2, and the settling time of 0.010 seconds, we can calculate the
required system bandwidth from the relation in equation \ref{eq7}:

$\omega_{n} = \frac{6.725}{\zeta T_{s}} = 820$ rad/sec, where $\zeta = 0.82$, and $T_{s} = 0.010$.

\subsection{PID Control Adventures} 
We have implemented a PID controller for the mirror, relying on trial and error tuning. To date, we have been unable to create a controller that fully meets the specifications for all cases.
We have chosen to move on to simulating and testing digital controllers synthesized from the full-order model. 
Using the plant model, a controller can be designed that is optimal for a 5th order system. This
controller is tabulated in [\cite{Dorf}]. In order to drive a steady-state error to zero, we need an integral term.
In order to stabilize the plant, we need the derivative term. So, a full-blown PID controller is
needed. The transfer function of the PID controller is:

\begin{equation}
C(s) = \frac{(K_{3} s^{2} + K_{2} s + K_{1})} {s}, 
  \label{eq8}
\end{equation}
where "s" is the Laplace transform variable, $K_{3}$ is the derivative gain, $ K_{2}$ is the proportional gain, and $K_{1}$ is the integral gain. Combining this transfer function with the plant, and closing the loop, we end up with the
closed-loop transfer function:

\begin{equation}
T(s) = \frac{C(s) P(s)}{1 + C(s) P(s)}, 
  \label{eq9}
\end{equation}

\begin{equation}
T(s) = \frac{\frac{(K_{3} s^{2} + K_{2} s + K_{1})}{s} (717.7s^{2} + 386.1 s + 5.272 10^{4})}{1 + \frac{(K_{3} s^{2} + K_{2} s + K_{1})} {s} (717.7 s^{2} + 386.1 s + 5.272 10^{4})}.
  \label{eq10}
\end{equation}

\subsection{Feed Forward} 
We can generate position, velocity and acceleration target functions from the real-time module front panel, and choose between a sinusoidal sigmoid, a Gompertz function sigmoid shape, or a "bang-bang" square shape for these signals.
The Gompertz function $ y(t) = a\exp(b\exp(ct)) $, plotted in Fig. 5, has been chosen for the ability to set the peak amplitude of the acceleration, and hence the current, through the coefficients $b$ and $c$.

\begin{figure}[H]
   \begin{center}
   \begin{tabular}{l r}
   \includegraphics[height=7cm]{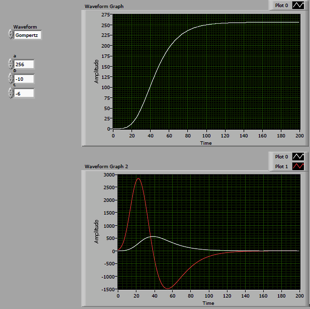} & \includegraphics[height=7cm]{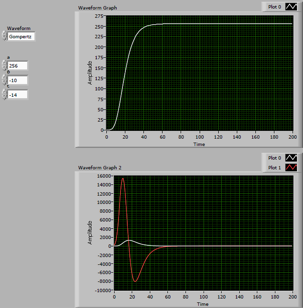} \\
   \end{tabular}
   \end{center}
      \caption[example] 
   { \label{fig:5} 
Two Gompertz functions (top panels: position, bottom panels: velocity and acceleration). Left panels are for b=-10, c=-6, right panels are for b=-10, c=-14.}
   \end{figure} 

\subsection{Other Strategies}
Since the major problem to overcome for an acceptable nutator performance is the switching and settling time, a time optimal controller should be considered.  The time optimal servo is implemented using a bang-bang servo, that is, the drive power is either fully positive, off, or fully negative.  This theoretically optimal system is clearly unworkable for our physical system.  Two other nearly time-optimal servos, the proximate time optimal servo (PTOS) \cite{adaptative}and the extended proximate time optimal servo (XPTOS) \cite{servo} are being investigated for implementation on the nutator.  The PTOS servo adds a linear band to the bang-bang control to avoid the actuator chattering that occurs from bang-bang switching when the error is fluctuating around zero.  XPTOS adds the ability to augment the controller with additional states to damp structural modes.  The nutator, in spite of the strengthening of the mirror structure, has several modes that can be damped using an advanced control algorithm.


\section{CURRENT PERFORMANCE} \label{sec:misc}

ALMA has set very stringent requirements on the nutator performance, with two main 
parameters that have been checked during the design modifications: (1) the pointing accuracy,
specified as a 1.3 arcsec RMS angular position error, and (2) the settling time between
on- and off-positions, constrained to 10 ms.  Identical specifications hold for throw angles
of up to 18 arcmin. The RMS pointing accuracy is evaluated discarding the 10 ms settling
time where overshooting and oscillatory features are tolerated. The 1.3 arcsec nutator
accuracy corresponds to a 0.3 arcsec pointing accuracy on the sky.
The current performance tests are limited to laboratory measurements, without any 
wind load or environmental influence.  
In a typical set-up, the nutator is alternating between an on-source and an off-source
position with variable throw angles, ranging from 60 arcsec to 18 arcmin. After initial 
single throws, where the basic functionality was optimized, we are now increasing the number
of cycles in order to assess the repeatability of the system. Repeatability is an additional 
key parameter for the nutator performance, as the number of duty cycles during one 
night can reach $\sim 10,000$ or more. This can then cumulate to millions of
cycles in a year time scale. 

\subsection{Single Cycles}

Table 2 illustrates typical single-cycle results.  Encoder-measured positions are recorded
every 60 $\mu$s. The results are for a  $0$ $\rightarrow$ $\theta$ $\rightarrow$ $0$  square cycles (see a typical cycle in Fig. 6) for ten different throw angles (+100 arcsec, +360 arcsec, +540 arcsec, +720 arcsec, +1080, and the opposite angles) and four frequencies 1 Hz, 2Hz, 5 Hz and 10 Hz, respectively. 
The RMS values are computed after a 10 ms throw time, until the start of the opposite throw. Grayed RMS values are for throws that are outside of the 1.3 arcsec RMS specification.
 
  \begin{figure}[H]
   \begin{center}
   \begin{tabular}{c}
   \includegraphics[height=4cm]{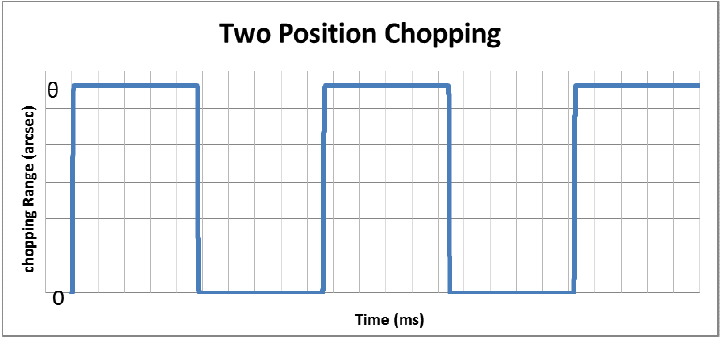}
   \end{tabular}
   \end{center}
   \caption[example] 
   { \label{repeatability} 
Two positions ($0$ $\rightarrow$ $\theta$ $\rightarrow$ $0$) nutating scheme.}
   \end{figure} 

\begin{table}[H]
\caption{Settling times and RMS errors for different chopping ranges and frequencies.} 
\label{tab:fonts}
\begin{center}       
\begin{tabular}{|p{1.5cm}|p{1.5cm}|p{2cm}|p{2.3cm}|p{2.3cm}|p{2.3cm}|p{2.3cm}|} 
\hline
\rule[-1ex]{0pt}{3.5ex} \bf Angle $\theta$ (arcsec) & \bf Frequency (Hz) & \bf Settling Time (ms) & \bf Error (RMS) at $+\theta$ position & \bf Error (RMS) at $+0$ position & \bf Error (RMS) at $-\theta$ position & \bf Error (RMS) at $-0$ position \\
\hline
\hline
\rule[-1ex]{0pt}{3.5ex}  100 & 1 & 10 & 0.18 & 0.30 & 0.68 & 0.23   \\
\hline
\rule[-1ex]{0pt}{3.5ex}  " & 2 & 10 & 0.24 & 0.44 & 0.58 & 0.21   \\
\hline
\rule[-1ex]{0pt}{3.5ex}  " & 5 & 10 & 0.28 & 0.37 & 0.77 & 0.34   \\
\hline
\rule[-1ex]{0pt}{3.5ex}  " & 10 & 10 & 0.55 & 0.72 & 1.13 & 0.46   \\
\hline
\rule[-1ex]{0pt}{3.5ex}  360 & 1 & 10 & 0.53 & 0.60 & 0.37 & 0.52   \\
\hline
\rule[-1ex]{0pt}{3.5ex}  " & 2 & 10 & 0.84 & 0.82 & 0.55 & 0.76   \\
\hline
\rule[-1ex]{0pt}{3.5ex}  " & 5 & 10 & 1.19 & 1.18 & 0.85 & 1.14   \\
\hline
\rule[-1ex]{0pt}{3.5ex}  " & 10 & 10 & \cellcolor[gray]{0.8} 1.71 & \cellcolor[gray]{0.8} 1.42 & 1.26 &  \cellcolor[gray]{0.8} 1.59 \\
\hline
\rule[-1ex]{0pt}{3.5ex}  540 & 1 & 10 & 0.46 & 0.54 & 0.43 & 0.43   \\
\hline
\rule[-1ex]{0pt}{3.5ex}  " & 2 & 10 & 0.78 & 0.71 & 0.64 & 0.70   \\
\hline
\rule[-1ex]{0pt}{3.5ex}  " & 5 & 10 & 1.12 & 0.89 & 0.96 & 0.91   \\
\hline
\rule[-1ex]{0pt}{3.5ex}  " & 10 & 10 & \cellcolor[gray]{0.8} 1.64 & 1.21 & \cellcolor[gray]{0.8} 1.41 & 1.29   \\
\hline
\rule[-1ex]{0pt}{3.5ex}  720 & 1 & 10 & 0.70 & 0.75 & 0.56 & 0.61   \\
\hline
\rule[-1ex]{0pt}{3.5ex}  " & 2 & 10 & 1.09 & 1.17 & 0.84 & 0.93   \\
\hline
\rule[-1ex]{0pt}{3.5ex}  " & 5 & 10 & 1.30 & \cellcolor[gray]{0.8} 1.43 & 1.21 & 1.23   \\
\hline
\rule[-1ex]{0pt}{3.5ex}  " & 10 & 10 & \cellcolor[gray]{0.8} 1.67 & \cellcolor[gray]{0.8} 1.89 & \cellcolor[gray]{0.8} 1.75 & \cellcolor[gray]{0.8} 1.60   \\
\hline
\rule[-1ex]{0pt}{3.5ex}  1080 & 1 & 10 & 0.82 & 0.93 & 0.91 & 0.90   \\
\hline
\rule[-1ex]{0pt}{3.5ex}  " & 2 & 10 & \cellcolor[gray]{0.8} 1.42 & \cellcolor[gray]{0.8} 1.43 & \cellcolor[gray]{0.8} 1.47 & \cellcolor[gray]{0.8} 1.41   \\
\hline
\rule[-1ex]{0pt}{3.5ex}  " & 5 & 10 & \cellcolor[gray]{0.8} 1.52 & \cellcolor[gray]{0.8} 1.63 & \cellcolor[gray]{0.8} 1.75 & \cellcolor[gray]{0.8} 1.53   \\
\hline
\rule[-1ex]{0pt}{3.5ex}  " & 10 & 10 & \cellcolor[gray]{0.8} 1.94 & \cellcolor[gray]{0.8} 2.25 & \cellcolor[gray]{0.8} 2.56 & \cellcolor[gray]{0.8} 2.07   \\
\hline
\end{tabular}
\end{center}
\end{table}

\subsection{Multiple Positions Switching} 
Here, the throw pattern is for $0$ $\rightarrow$ $+\theta$ $\rightarrow$ $0$ $\rightarrow$ $-\theta$ cycles (see a typical cycle in Fig. 7). Again, the grayed combinations of angle throws and
frequencies are the ones outside of the specifications. 

  \begin{figure}[H]
   \begin{center}
   \begin{tabular}{c}
   \includegraphics[height=4cm]{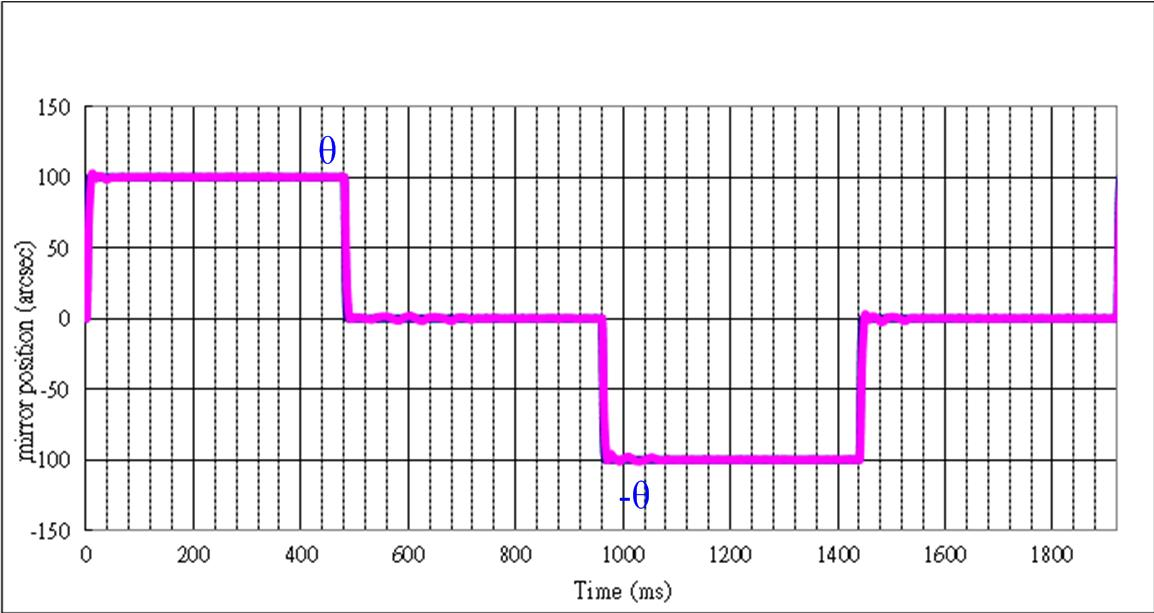}
   \end{tabular}
   \end{center}
   \caption[example] 
   { \label{repeatability} 
Multiple positions ($0$ $\rightarrow$ $+\theta$ $\rightarrow$ $0$ $\rightarrow$ $-\theta$) nutating scheme.}
   \end{figure}

\begin{table}[H]
\caption{Settling times and RMS errors for different chopping ranges and frequencies.} 
\label{tab:fonts}
\begin{center}       
\begin{tabular}{|p{1.5cm}|p{1.5cm}|p{2.3cm}|p{2.3cm}|p{2.3cm}|p{2.3cm}|} 
\hline
\rule[-1ex]{0pt}{3.5ex} \bf Angle $\theta$ (arcsec) & \bf Frequency (Hz)  & \bf Error (RMS) at $-0$ position & \bf Error (RMS) at $+\theta$ position & \bf Error (RMS) at $+0$ position & \bf Error (RMS) at $-\theta$ position  \\
\hline
\hline
\rule[-1ex]{0pt}{3.5ex}  100 & 1  & 0.25 & 0.27 & 0.92 & 0.37   \\
\hline
\rule[-1ex]{0pt}{3.5ex}  " & 2  & 0.31 & 0.36 & 0.98 & 0.37   \\
\hline
\rule[-1ex]{0pt}{3.5ex}  " & 5  & 0.53 & 0.52 & 1.12 & 0.68   \\
\hline
\rule[-1ex]{0pt}{3.5ex}  " & 10  & 0.66 & 0.55 & 0.88 & 0.66   \\
\hline
\rule[-1ex]{0pt}{3.5ex}  360 & 1  & 1.06 & 0.78 & 0.44 & 0.77   \\
\hline
\rule[-1ex]{0pt}{3.5ex}  " & 2  & \cellcolor[gray]{0.8} 1.33 & 1.07 & 0.58 & 0.91   \\
\hline
\rule[-1ex]{0pt}{3.5ex}  " & 5  & \cellcolor[gray]{0.8} 2.30 & 0.68 & \cellcolor[gray]{0.8} 1.71 & \cellcolor[gray]{0.8} 2.30   \\
\hline
\rule[-1ex]{0pt}{3.5ex}  " & 10  & \cellcolor[gray]{0.8} 2.53 & \cellcolor[gray]{0.8} 1.99 & 0.95 &  \cellcolor[gray]{0.8} 1.78 \\
\hline
\rule[-1ex]{0pt}{3.5ex}  540 & 1  & 0.81 & 0.58 & 0.52 & 0.41   \\
\hline
\rule[-1ex]{0pt}{3.5ex}  " & 2  & 1.10 & 0.88 & 0.55 & 0.51   \\
\hline
\rule[-1ex]{0pt}{3.5ex}  " & 5  & \cellcolor[gray]{0.8} 1.65 & \cellcolor[gray]{0.8} 1.38 & 0.85 & 0.87   \\
\hline
\rule[-1ex]{0pt}{3.5ex}  " & 10  & \cellcolor[gray]{0.8} 2.20 & \cellcolor[gray]{0.8} 1.38 & \cellcolor[gray]{0.8} 1.41 & \cellcolor[gray]{0.8} 1.31   \\
\hline
\rule[-1ex]{0pt}{3.5ex}  720 & 1  & 0.67 & 0.68 & 0.45 & 0.58   \\
\hline
\rule[-1ex]{0pt}{3.5ex}  " & 2  & 0.90 & 1.05 & 0.66 & 0.84   \\
\hline
\rule[-1ex]{0pt}{3.5ex}  " & 5  & 0.99 & \cellcolor[gray]{0.8} 1.70 & \cellcolor[gray]{0.8} 1.53 & \cellcolor[gray]{0.8} 1.37   \\
\hline
\rule[-1ex]{0pt}{3.5ex}  " & 10  & \cellcolor[gray]{0.8} 1.85 &  1.14 &  1.29 & \cellcolor[gray]{0.8} 1.59   \\
\hline
\rule[-1ex]{0pt}{3.5ex}  1080 & 1  & 0.98 & 0.90 & 0.85 & 1.15   \\
\hline
\rule[-1ex]{0pt}{3.5ex}  " & 2  & \cellcolor[gray]{0.8} 1.64 & \cellcolor[gray]{0.8} 1.48 &  1.03 & \cellcolor[gray]{0.8} 1.67   \\
\hline
\rule[-1ex]{0pt}{3.5ex}  " & 5  & \cellcolor[gray]{0.8} 2.62 & \cellcolor[gray]{0.8} 2.07 & \cellcolor[gray]{0.8} 2.39 & \cellcolor[gray]{0.8} 2.82   \\
\hline
\rule[-1ex]{0pt}{3.5ex}  " & 10  & \cellcolor[gray]{0.8} 3.03 & \cellcolor[gray]{0.8} 1.52 & \cellcolor[gray]{0.8} 1.97 & \cellcolor[gray]{0.8} 1.93   \\
\hline
\end{tabular}
\end{center}
\end{table}    
   
\subsection{Repeatability}

   \begin{figure}
   \begin{center}
   \begin{tabular}{c}
   \includegraphics[scale=0.5]{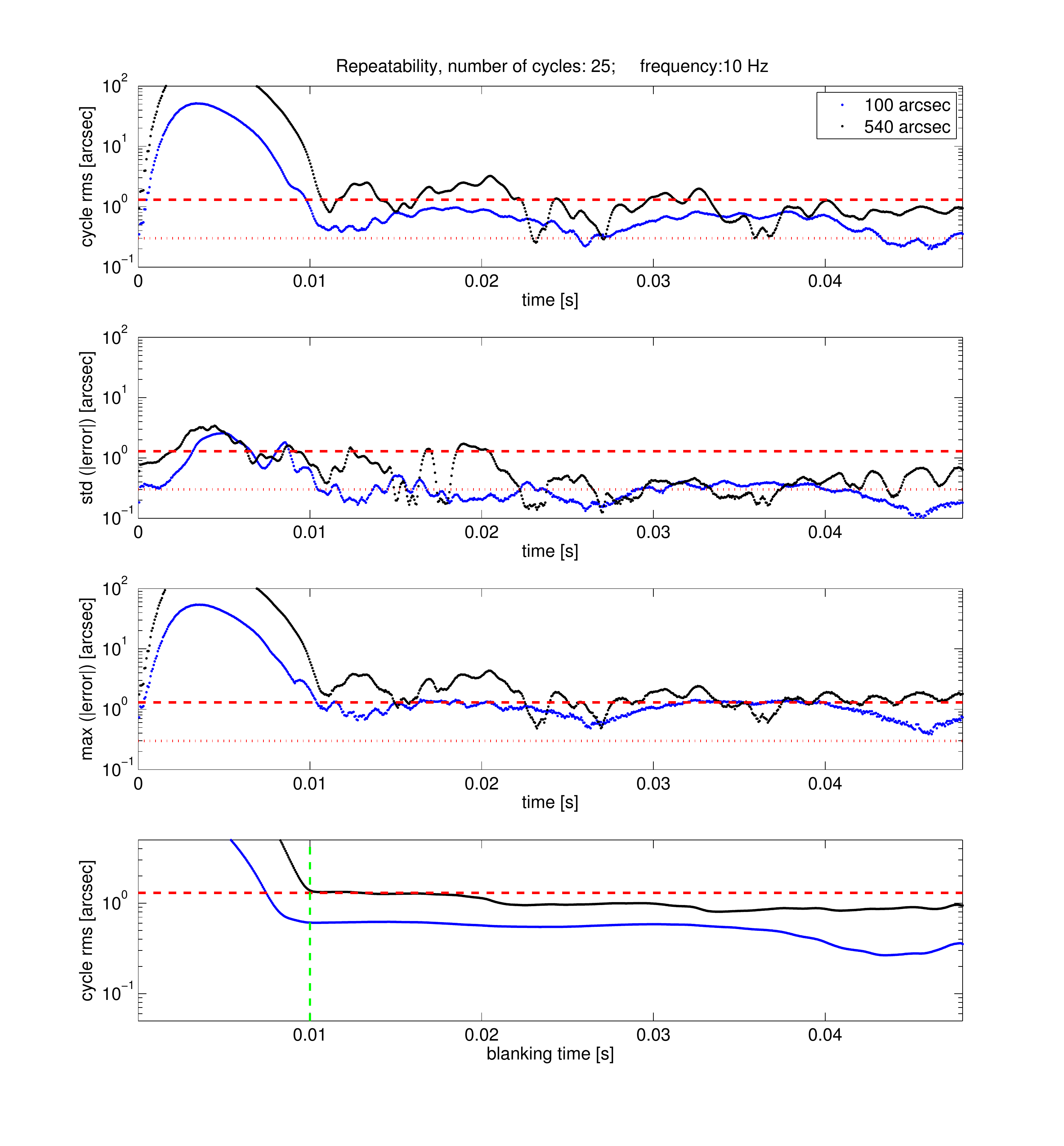}
   \end{tabular}
   \end{center}
   \caption[example] 
   { \label{repeatability} 
Repeatability results for 25 cycles for 100 arcsec (blue) and 540 arcsec (black) throws. The dashed line is the 1.3 arcsec  ALMA pointing specification, the dotted line marks 0.3 arcsec for reference.}
   \end{figure} 

Results for first repeatability tests are shown in Fig.\ref{repeatability} for 25 cycles
(25 consecutive on-/off-source positions). Overlaid are the data for 100 arcsec and 540 arcsec
throws at 10 Hz. Data are again recorded every 60 $\mu$s. In order to evaluate the repeatability, 
all the cycles are exactly aligned in time; i.e., at every time step $t_i$ in a cycle, the recorded
positions of all the cycles are compared. The sequence of commanded positions is always the 
same in every cycle. The top panel in  Figure \ref{repeatability} shows the RMS deviations
(commanded versus recorded positions) over the 25 cycles as a function of time. A clear peak
during the settling time is apparent before the RMS values remain within a smaller range. 
The second and third panels show standard and maximum deviations at every time step. 
The standard deviations are typically smaller than or comparable to the RMS values, which 
indicates that there is no severe spread in the deviations. Similarly, the maximum values do not
show any significant outliers. This leads to the conclusion that the point-to-point repeatability
over the 25 cycles is very good. Nevertheless, it has to be remarked that the similar features
in the maximum and RMS curves mean that single maximum deviations can drive the cycle 
RMS values. However, this is not of a concern as along as the RMS values are within the 
specified range of 1.3 arcsec (red dashed lines). \\
We choose to define the repeatability as the RMS value of the entire cycle RMS (top panel in 
Figure \ref{repeatability}). The bottom panel in Figure \ref{repeatability} displays this value
as a function of blanking time. Here, an increasingly larger time interval, starting from 0, 
is discarded from the top panel when calculating the RMS value. It is clearly visible that the 
repeatability converges to a plateau after discarding the initial settling time. This approach is
further instructive if an additional blanking time (beyond the 10 ms settling time) needs to 
be defined to bring the performance into the desired range.  For the examples shown in 
Figure \ref{repeatability}, this seems irrelevant as the repeatability converges to a value 
below or at 1.3 arcsec within 10 ms.

\section{CONCLUSION} \label{sec:misc}
While there are still possibilities for improvement, mostly in the software area, the ALMA nutator is now mature enough for 
factory acceptance test, as well as site testing which is expected to take place before the end of 2012.

\appendix    

%
\
\bibliography{report}   
\bibliographystyle{spiebib}   

\end{document}